\documentclass[conference]{IEEEtran}
\IEEEoverridecommandlockouts
\usepackage{cite}
\usepackage{amsmath,amssymb,amsfonts}
\usepackage{algorithmic}
\usepackage{graphicx}
\usepackage{textcomp}
\usepackage{xcolor}
\usepackage{booktabs}
\usepackage{tabularx}
\usepackage{threeparttable}
\usepackage{svg}
\usepackage{subcaption}
\usepackage{rotating}
\usepackage{xspace}
\PassOptionsToPackage{hyphens}{url}\usepackage{hyperref}

\usepackage[frozencache,cachedir=.]{minted}
\def\BibTeX{{\rm B\kern-.05em{\sc i\kern-.025em b}\kern-.08em
    T\kern-.1667em\lower.7ex\hbox{E}\kern-.125emX}}

\def\bf2{{\mbox{BlueField-2}}}
\newcommand{\pktgen}{\texttt{pktgen}\xspace}
\newcommand{\stressng}{\texttt{stress-ng}\xspace}

\begin{document}

\title{Performance Characteristics of the \mbox{BlueField-2 SmartNIC}\\
  \thanks{This material is based upon work supported by the U.S. Department of Energy, Office of Science, Office of Advanced Scientific Computing Research under Field Work Proposal Number 20-023266. Sandia National Laboratories is a multimission laboratory managed and operated by National Technology \& Engineering Solutions of Sandia, LLC., a wholly owned subsidiary of Honeywell International Inc., for the U.S. Department of Energy's National Nuclear Security Administration under contract DE-NA0003525. This paper describes objective technical results and analysis. Any subjective views or opinions that might be expressed in the paper do not necessarily represent the views of the U.S. Department of Energy or the United States Government.}
}

\author{\IEEEauthorblockN{Jianshen Liu}
\IEEEauthorblockA{\textit{UC Santa Cruz} \\
jliu120@ucsc.edu}
\and
\IEEEauthorblockN{Carlos Maltzahn}
\IEEEauthorblockA{\textit{UC Santa Cruz} \\
carlosm@ucsc.edu}
\and
\IEEEauthorblockN{Craig Ulmer}
\IEEEauthorblockA{\textit{Sandia National Laboratories} \\
cdulmer@sandia.gov}
\and
\IEEEauthorblockN{Matthew Leon Curry}
\IEEEauthorblockA{\textit{Sandia National Laboratories} \\
mlcurry@sandia.gov}
}

\maketitle

\begin{abstract}
  High-performance computing (HPC) researchers have long envisioned scenarios where application workflows could be improved through the use of programmable processing elements embedded in the network fabric. Recently, vendors have introduced programmable Smart Network Interface Cards (SmartNICs) that enable computations to be offloaded to the edge of the network. There is great interest in both the HPC and high-performance data analytics (HPDA) communities in understanding the roles these devices may play in the data paths of upcoming systems.
  
  This paper focuses on characterizing both the networking and computing aspects of NVIDIA's new \bf2 SmartNIC when used in a 100Gb/s Ethernet environment. For the networking evaluation we conducted multiple transfer experiments between processors located at the host, the SmartNIC, and a remote host. These tests illuminate how much effort is required to saturate the network and help estimate the processing headroom available on the SmartNIC during transfers. For the computing evaluation we used the \stressng benchmark to compare the \bf2 to other servers and place realistic bounds on the types of offload operations that are appropriate for the hardware.
  
  Our findings from this work indicate that while the \bf2 provides a flexible means of processing data at the network's edge, great care must be taken to not overwhelm the hardware. While the host can easily saturate the network link, the SmartNIC's embedded processors may not have enough computing resources to sustain more than half the expected bandwidth when using kernel-space packet processing. From a computational perspective, encryption operations, memory operations under contention, and on-card IPC operations on the SmartNIC perform significantly better than the general-purpose servers used for comparisons in our experiments. Therefore, applications that mainly focus on these operations may be good candidates for offloading to the SmartNIC.
\end{abstract}


\section{Introduction}

The rise in popularity of large-scale data analytics over the last decade has resulted in a \emph{deluge of data} for system architects to manage. Enterprises now aggressively collect data from nearly every nook and cranny of the world in hopes that powerful analytics will be able to harvest insights that were previously unseen. IDC recently reported that the compound annual growth rate of the Global StorageSphere is expected to be 20.4\%, reaching a total data size of 8.9ZB by 2024~\cite{b1}. While we are excited by the possibilities that this increase in analytics scale may provide, it is important to acknowledge that it is challenging for system architects to deliver platforms that can meet the growing storage and compute needs of this application space.

Traditional big-data systems have largely met performance goals through horizontal scaling: when additional storage or compute capabilities are needed, architects simply purchase new servers and add them to the end of the system. Given that data centers have finite space, cooling, and power, many researchers have recently turned to improving vertical scaling by leveraging high-performance computing (HPC) technologies and approaches in the platform. The resulting high-performance data analytics (HPDA) platforms feature high-speed networks with RDMA capabilities, computational accelerators such as GPUs and TPUs~\cite{b58}, and low-latency NVMe storage devices. While these devices greatly improve the computational density that can be achieved in HPDA systems, they also re-emphasize a fundamental scaling challenge hindering this field: modern analytics applications are often bottlenecked by the overhead of data movement~\cite{b57,b59,b60,b61}. Therefore, one key challenge to HPDA is minimizing the overhead of data movement between functions and hardware components.

A fundamental mechanism for reducing data movement in distributed systems is to place processing elements throughout the architecture to enable users to push computations to locations in the data path where filtering will have the most impact. While traditional big-data efforts have largely focused on using host processors to collocate computations and data, some commercial hardware vendors have made the embedded processors in their devices available for users to execute their own applications. This programmability enables researchers to consider an environment where operations can be placed at host systems, storage systems, storage devices, and at locations within the network fabric.

\subsection{SmartNICs}
The need for greater flexibility in network security and cloud infrastructure applications has motivated network vendors to develop Smart Network Interface Cards or \emph{SmartNICs}~\cite{b44}. In addition to providing the host with a high-speed connection to the network fabric, SmartNICs feature a variety of hardware resources (e.g., CPU cores, DRAM, FPGAs, and/or non-volatile storage) that can be leveraged by developers. While capabilities vary greatly between products, the primary benefit of this hardware is its \emph{programmability}: developers can embed application-specific functionality at the network's edge. There are multiple ways in which we expect SmartNICs will improve HPDA platforms:

\begin{itemize}
\item \textbf{Reduce host resource utilization:}
  End-to-end network communication protocols can consume substantial resources, especially at high speeds. For example, TCP-based packet processing is taxing at 100Gb/s rates as the protocol must properly buffer large amounts of in-transit data and react to asynchronous messages in a timely manner. Offloading the work to a network processor can reduce messaging overhead to the host and remove an important cause of cache pollution~\cite{b33,b34}.
  
\item \textbf{Accelerate specialized operations:}
  Some SmartNICs include specialized hardware that can perform common streaming operations (e.g., encryption and regular expression acceleration~\cite{b6,b41}) more efficiently than general-purpose processors. Domain-specific hardware has drawn increasing research interest in recent years~\cite{b35,b36,b37,b38,b39}, and has been considered an important source of computing power to compensate the need of surpassing performance in the age of post-Moore's Law~\cite{b38,b39,b29,b31}.
  
\item \textbf{Improve power efficiency:}
  The ARM and FPGA computing resources typically found in SmartNICs have been demonstrated to be \textit{more power-efficient}~\cite{b3,b4,b40,b41,b42} (therefore more cost-efficient as well) than the corresponding components from hosts on various tasks (e.g., real-time analytics~\cite{b5}, distributed transactions, and replicated key-value stores~\cite{b2}), while still being able to deliver similar or better performance.

\end{itemize}

\subsection{NVIDIA \bf2 SmartNIC Specifications}
\label{sec:bf2_spec}
NVIDIA's recent \bf2 card provides an example of the capabilities found in a current SmartNIC card. As listed in Table~\ref{table:bluefield-2_features}, the \bf2 card supports two high-speed network interfaces and features a multicore ARM processor that runs a customized version of Linux. The card can be configured to operate in one of three distinct modes at boot time: separated host mode, embedded function mode, and restricted mode~\cite{b16}. In separated host mode the host and SmartNIC processors function as independently-addressed network entities that share the same network interface. In embedded function mode all network traffic to and from the host interface is always routed through the embedded ARM processor. This mode is typically used to implement a virtual network switch on the ARM that can filter, steer, monitor, and apply QoS rules to packets traveling between the host and network. The restricted mode is an extension to the embedded function mode that puts additional limitations on the host side for security purposes. In this paper we primarily focus on separated host mode because it provides a straightforward environment for offloading existing services to the SmartNIC. Additional experiments involving embedded function mode are presented to observe the benefits of performing computations on network flows.

The experiments described in this paper were conducted on NVIDIA \bf2 SmartNICs available in the National Science Foundation's CloudLab~\cite{b9}. Cards were loaded with the Ubuntu 20.04 variant of the 3.5.1.11601 release of the NVIDIA BlueField software stack. The network cards were connected to a 100Gb/s Ethernet network in the cloud.

\begin{table}
\begin{threeparttable}
\caption{Main Features of the \bf2 SmartNIC}
\label{table:bluefield-2_features}
\begin{tabularx}{\linewidth}{ l X }
\toprule
\textbf{CPU} & \begin{tabular}[c]{@{}l@{}}8 ARMv8 A72 cores (64-bit) @ 2.5 GHz,\\1MB L2 cache per 2 cores, 6MB L3 cache\end{tabular} \\
\midrule
\textbf{DRAM} & 16 GB on-board DDR4-1600 \\ 
\midrule
\textbf{Storage} & eMMC flash memory \\ 
\midrule
\textbf{Network} & \begin{tabular}[c]{@{}l@{}}Ethernet or InfiniBand:\\dual ports of 10/25/50/100 Gb/s,\\or a single port of 200 Gb/s\end{tabular} \\ 
\midrule
\textbf{Accelerators} & \begin{tabular}[c]{@{}l@{}}\begin{tabular}{@{\labelitemi\hspace{\dimexpr\labelsep+0.5\tabcolsep}}l}Hardware root of trust\\RegEx\\IPsec/TLS data-in-motion encryption\\AES-XTS 256/512-bit data-at-rest encryption\\SHA 256-bit hardware acceleration\\Hardware public key accelerator\\True random number generator\end{tabular}\end{tabular} \\ 
\midrule
\textbf{PCIe} & Gen 4.0 x16 \\
\midrule
\textbf{OS} & Ubuntu 20.04 (kernel 5.4.0-1007-bluefield) \\
\bottomrule
\end{tabularx}
\begin{tablenotes}\scriptsize
\item All information is available on~\cite{b10}.
\end{tablenotes}
\end{threeparttable}
\end{table}

\vspace{2mm}

\subsection{SmartNIC Research Direction}

While SmartNICs such as the \bf2 offer a great deal of opportunity for improving data center workloads, there are a number of research challenges that must be addressed in designing a production system that dynamically offloads computation into the network fabric. These challenges relate to the \textit{what}, \textit{when}, and \textit{how} of the system:

\begin{itemize}
\item \textbf{What}: The \emph{characterization problem} focuses on understanding what functions should be offloaded to a given SmartNIC indicated by the performance characterization of the hardware. Identifying these functions is challenging because the variable space for this problem is inherently large, which forces us to consider heterogeneous resources, complex code paths, and dynamic workloads. Most existing research tackles this challenge in specific domains~\cite{b3,b4,b2,b5}, but still, no general solutions have been carried out.

\item \textbf{When}: The \emph{scheduling problem} involves making strategic decisions about when to offload computations to different processing elements that are available along a workflow's data path. Scheduling is challenging because it must take into account both the dynamic load conditions of the platform and the impact placement may have on a workflow. For example, it may be beneficial to perform filtering operations as close to storage as possible to reduce data, but pressure from other workflows on the storage system may necessitate deferring the operation until later in the data path. 

\item \textbf{How}: Finally, the \emph{architecture problem} studies how the overall system should be designed to maximize benefit. This problem space focuses on improving the usability, flexibility, and security of the system. Services developed in this work will provide workflows with runtime information that improves scheduling decisions, and ensure fairness in platforms with concurrent workflows.
  
\end{itemize}

In this paper, we focus on tackling the first problem described above by characterizing the performance capability of a representative SmartNIC. Specifically, we want to answer the following two questions:

\begin{enumerate}
\item Given that the main function of a network adapter is to send and receive packets, for SmartNICs, what is the maximum available processing headroom for applications when transmitting a given batch of packets?
\item What types of operations are profitable to perform on the SmartNIC in comparison to the host?
\end{enumerate}

Understanding how well a SmartNIC can perform computations under duress is essential for determining what work can be offloaded to the network. To answer the first question, we use Linux's \pktgen~\cite{b11,b12} traffic generation tool to determine how much delay between packet transmissions can be tolerated before network throughput drops. This delay gives an estimate of how much free time a processor has for performing offloaded computations. For the second question, we use the \stressng~\cite{b20} tool to observe how well the SmartNIC performs a number of targeted computational tasks. Comparing the \bf2 SmartNIC's test results with those from other hosts helps reveal the underlying characteristics of the hardware and identify operations that may be profitable for the SmartNIC to perform.

The rest of this paper is structured as follows. Section~\ref{sec:processing_headroom} provides an overview of \pktgen and explains how we use it to estimate processing headroom. We report performance measurements for multiple configurations and then analyze the results. Section~\ref{sec:profitable_operations} describes our motivation for using \stressng, the systems we used for comparison, and the results of the performance experiments. Finally, we conclude the paper with a summary of the findings and insights we learned during this effort in section~\ref{sec:conclusion}.

\section{Characterizing Processing Headroom During Network Transfers}
\label{sec:processing_headroom}

\subsection{Benchmark Considerations}
The goal of our first performance evaluation is to determine how much processing headroom is available on the \bf2 card's embedded processors when the SmartNIC is busy transmitting network data at maximum speed. For this experiment we run a packet generator on the SmartNIC and then insert a variable amount of delay between packet bursts to model a scenario where computations are applied to a network stream. Finding the maximum amount of delay that can be tolerated before the network's performance drops gives an estimate of the amount of time the hardware has available for offloading computations. Given that we are seeking an upper bound on this time, it is useful to select a packet generator that yields maximum network bandwidth with minimal overhead. After our initial experiments with user space utilities such as iPerf~\cite{b13}, nuttcp~\cite{b52}, and Netperf~\cite{b14} yielded suboptimal performance, we transitioned to the Linux \pktgen tool. \pktgen is a kernel space tool that injects UDP packets directly into the kernel IP network stack.

There are multiple benefits to using \pktgen in this evaluation. First, we observed that its performance was roughly 15\% higher than the aforementioned user space benchmark utilities in resource-restricted environments. Second, \pktgen has built-in support for symmetric multiprocessing and can start a kernel thread on each CPU core. This feature is important in 100Gb/s Ethernet environments where it is nearly impossible to saturate the network with a single core (e.g., a single-instance of iPerf achieved less than 40Gb/s in a previous study~\cite{b15}). Finally, \pktgen provides multiqueue support that allows a socket buffer's transmission queue to be mapped to the running thread's CPU core. This optimization reduces the overhead of cross-core communication and improves throughput.

There are three main options that we supplied to \pktgen to adjust the settings of our experiments. First, the ``delay'' option was used to regulate the minimum amount of time allocated to send a batch (or burst) of packets. Varying this option provides us with a way to determine how much additional processing the SmartNIC can perform before network throughput drops. Second, the ``clone\_skb'' option controls how often a network packet is reused for transmission. Setting this option to zero removes memory allocation times. Future tests may use larger values to increase allocation costs for communication. Finally, the ``burst'' option specifies the number of packets queued up before enabling the bottom half of the network stack. This option can be used to adjust interrupt coalescing.

\subsection{Experiment Setup and Results}
\pktgen threads use an infinite loop model such that when a \pktgen thread is enabled, the associated CPU core will be fully occupied. Therefore, measuring the processing headroom with this tool requires two steps. In the first step, we need to measure the minimum configuration with which the SmartNIC can achieve the highest possible bandwidth. To be specific, we set the ``clone\_skb'' to 0, and gradually increased the number of threads and the value of ``burst'' while recording the throughput changes. Once we have found the minimum configuration, the second step is to modify the ``delay'' setting to inject an artificial delay for each burst of packets. We find the maximum delay the SmartNIC can withstand before throughput drops for a particular number of threads. Thus, the maximum processing headroom available when transmitting a given batch of packets can be calculated by subtracting the time spent on sending the batch of packets without delay evaluated in the first step from the maximum delay evaluated in this step. Note that we kept the default MTU value (1500B) in all experiments as this should be the most common scenario. The experiments were conducted with the \bf2 card in both the separated host mode and the embedded function mode.

\subsubsection{Evaluation in the Separated Host Mode}
\begin{figure*}[!ht]
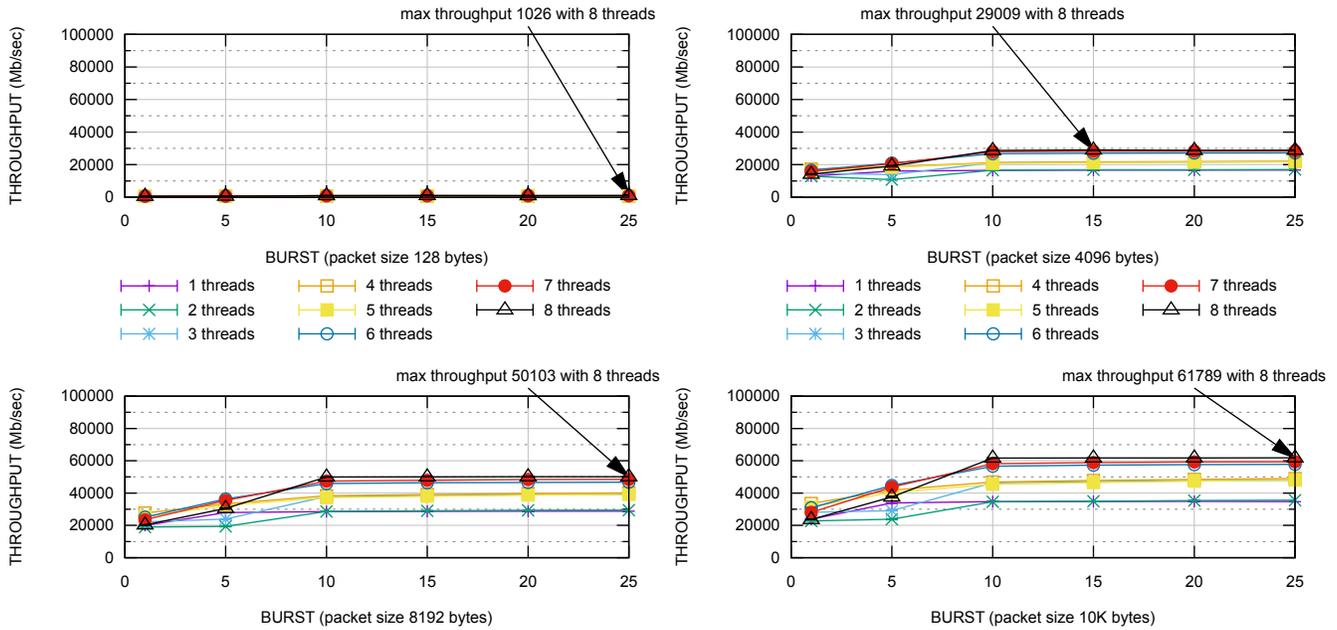

  \begin{subfigure}{\columnwidth}
    \label{fig:smartnic_pkt_size_128bytes}
    \includesvg[inkscapelatex=false, width=\columnwidth]{images/smartnic_pkt_size_128bytes.svg}
  \end{subfigure}%
  \begin{subfigure}{\columnwidth}
    \label{fig:smartnic_pkt_size_4096bytes}
    \includesvg[inkscapelatex=false, width=\columnwidth]{images/smartnic_pkt_size_4096bytes.svg}
  \end{subfigure}
  \begin{subfigure}{\columnwidth}
    \label{fig:smartnic_pkt_size_8192bytes}
    \includesvg[inkscapelatex=false, width=\columnwidth]{images/smartnic_pkt_size_8192bytes.svg}
  \end{subfigure}%
  \begin{subfigure}{\columnwidth}
    \label{fig:smartnic_pkt_size_10000bytes}
    \includesvg[inkscapelatex=false, width=\columnwidth]{images/smartnic_pkt_size_10000bytes.svg}
  \end{subfigure}
  \caption{Throughput results from the \bf2 SmartNIC in host separated mode}
  \label{fig:pktgen_smartnic_step_1}
\end{figure*}

Conducting the first step of the experiment involved performing a sweep of packet sizes to find the minimum configuration settings that would generate the maximum bandwidth. Packet sizes ranged from 128B to 10KB. Packets larger than exactly 10KB caused the test process to crash. Throughput measurements are presented in Figure~\ref{fig:pktgen_smartnic_step_1}.

The results for the \bf2 surprised us. While we knew that the \bf2 card's embedded processors would have trouble saturating the 100Gb/s network link, we expected that many worker threads generating large packets would be able to fill a significant portion of the available bandwidth. At best we found the card could only generate approximately 60\% of the total bandwidth. On the one hand, this evaluation shows that the packet processing task itself is resource-intensive, even for the most advanced SmartNICs available nowadays. On the other hand, we see that offloading functions to this SmartNIC without changing the traditional network processing path used by these functions cannot achieve the best performance.

\begin{figure}[!ht]
  \includesvg[inkscapelatex=false, width=\columnwidth]{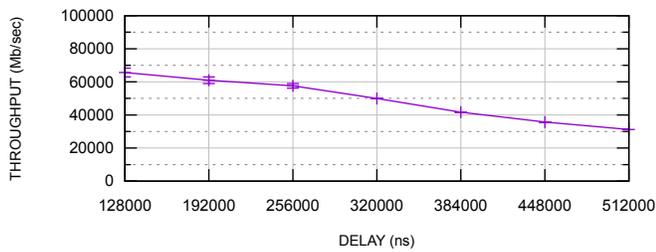}
  \caption{Throughput results from the \bf2 SmartNIC with different delay configurations (8 threads, packet size 10KB, burst 25)}
  \label{fig:smartnic_pkt_size_10000bytes_delays}
\end{figure}

After accepting that the \bf2 can only realistically support approximately 50 Gb/s of traffic, the next step in our evaluation was to measure how much delay can be added to packet generation to determine how much processing headroom exists for this data rate. As presented in Figure~\ref{fig:smartnic_pkt_size_10000bytes_delays} the maximum delay before bandwidth degradation is approximately 320$\mu$s. If we subtract the time required for each burst of transmission without delay from this delay, and convert the result to the available CPU percentage per core per burst of transmission (10KB x 25 = 250KB), we get 22.8\% of CPU time left for application logic on the ARM cores if we only aim to use up to 50\% of the full network bandwidth.

\begin{figure*}[!ht]
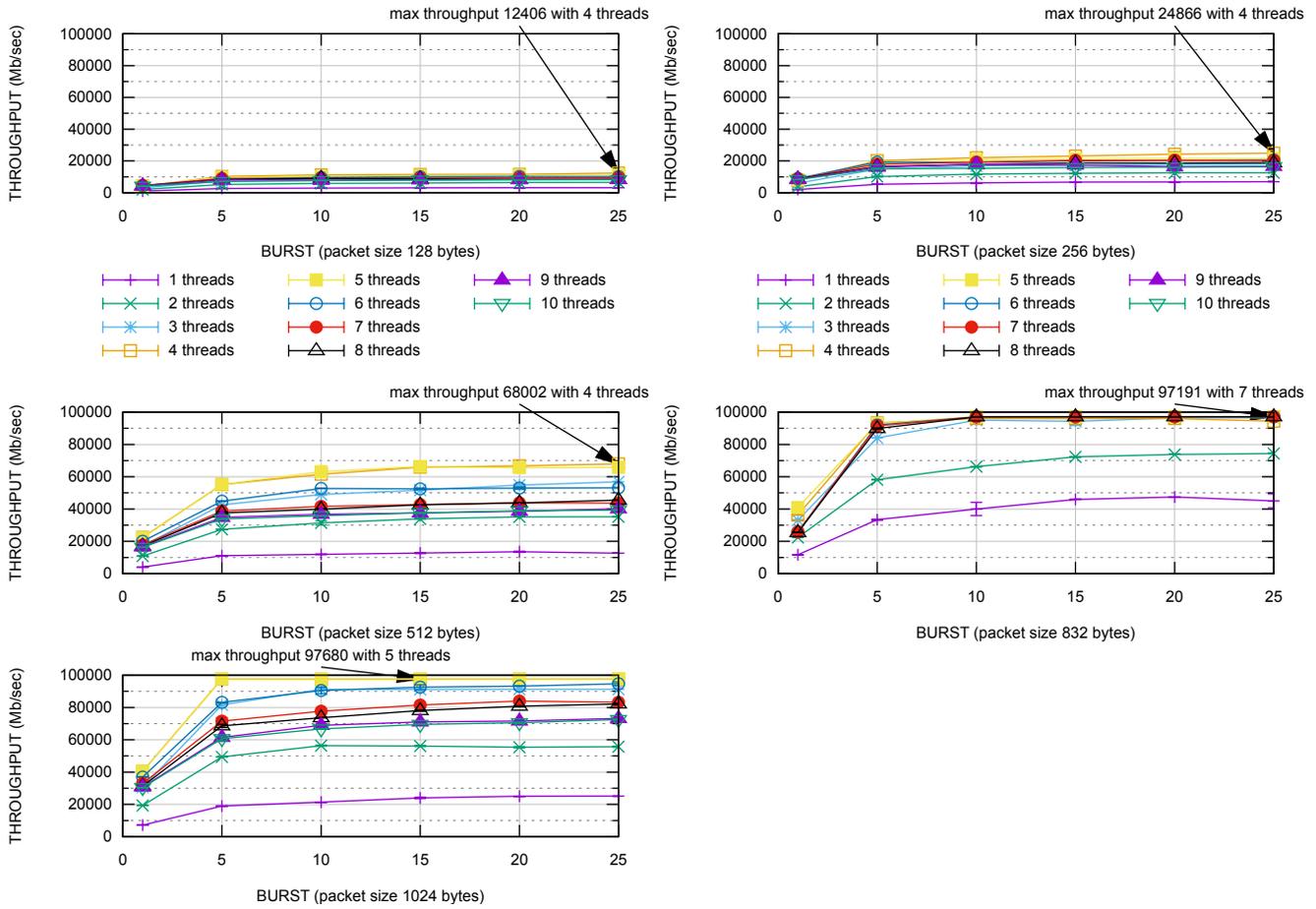

  \begin{subfigure}{\columnwidth}
    \label{fig:r7525_pkt_size_128bytes}
    \includesvg[inkscapelatex=false, width=\columnwidth]{images/r7525_pkt_size_128bytes.svg}
  \end{subfigure}%
  \begin{subfigure}{\columnwidth}
    \label{fig:r7525_pkt_size_256bytes}
    \includesvg[inkscapelatex=false, width=\columnwidth]{images/r7525_pkt_size_256bytes.svg}
  \end{subfigure}
  \begin{subfigure}{\columnwidth}
    \label{fig:r7525_pkt_size_512bytes}
    \includesvg[inkscapelatex=false, width=\columnwidth]{images/r7525_pkt_size_512bytes.svg}
  \end{subfigure}%
  \begin{subfigure}{\columnwidth}
    \label{fig:r7525_pkt_size_832bytes}
    \includesvg[inkscapelatex=false, width=\columnwidth]{images/r7525_pkt_size_832bytes.svg}
  \end{subfigure}
   \begin{subfigure}{\columnwidth}
    \label{fig:r7525_pkt_size_1024bytes}
    \includesvg[inkscapelatex=false, width=\columnwidth]{images/r7525_pkt_size_1024bytes.svg}
  \end{subfigure}
  \caption{Throughput results from the r7525 machine in host separated mode}
  \label{fig:pktgen_r7552_host_separated_mode_step_1}
\end{figure*}

For comparison with a general-purpose processor, we ran a similar set of tests on the host where the \bf2 card resides (CloudLab machine type r7525). The host has two 32-core AMD 7542 CPUs (2.9 GHz) and 512GB DDR4-3200 memory. We varied packet sizes from 128B to 1KB and saw that the 100Gb/s link could be saturated with a packet size of only 832B. The results are presented in Figure~\ref{fig:pktgen_r7552_host_separated_mode_step_1}. Thanks to the host's much more powerful system resources, saturating the network can be accomplished with only 5 threads (corresponding to 5 vCPU cores) and a burst size of 25. While additional threads had a minimal impact on throughput for the 832B packet case, we observed a performance drop in larger packet sizes such as 1KB when more threads were added. We believe this drop is due to resource contention.

\begin{figure}[!ht]
  \includesvg[inkscapelatex=false, width=\columnwidth]{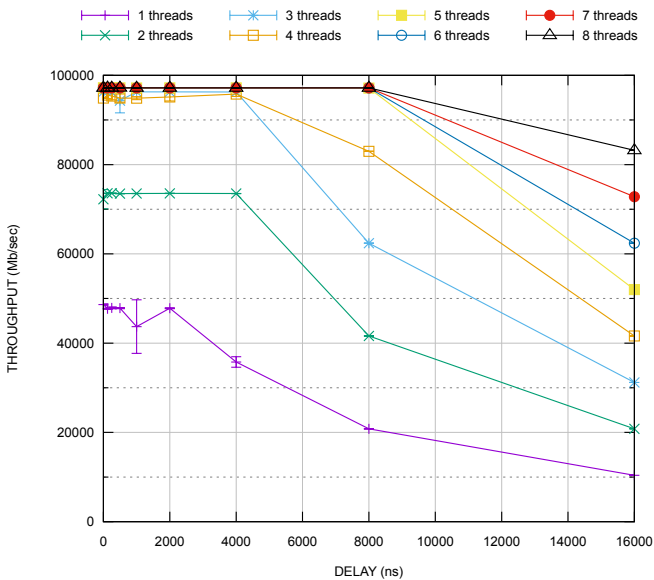}
  \caption{Throughput results from the r7525 machine with different delay configurations (packet size 832B, burst 25)}
  \label{fig:r7525_pkt_size_1024bytes_delays}
\end{figure}

Using the minimum configuration (5 threads and a burst of 25), we can then inject a small delay per burst to see how well the full bandwidth can sustain on the host (Figure~\ref{fig:r7525_pkt_size_1024bytes_delays}). Based on the results, the host can afford 8$\mu$s delay per burst without using additional threads. This delay is equivalent to $<$1\% of CPU time available for handling application logic on these five cores. However, this savings is not significant given that the host has an additional 123 vCPU cores available for use by applications!

\begin{figure}[!ht]
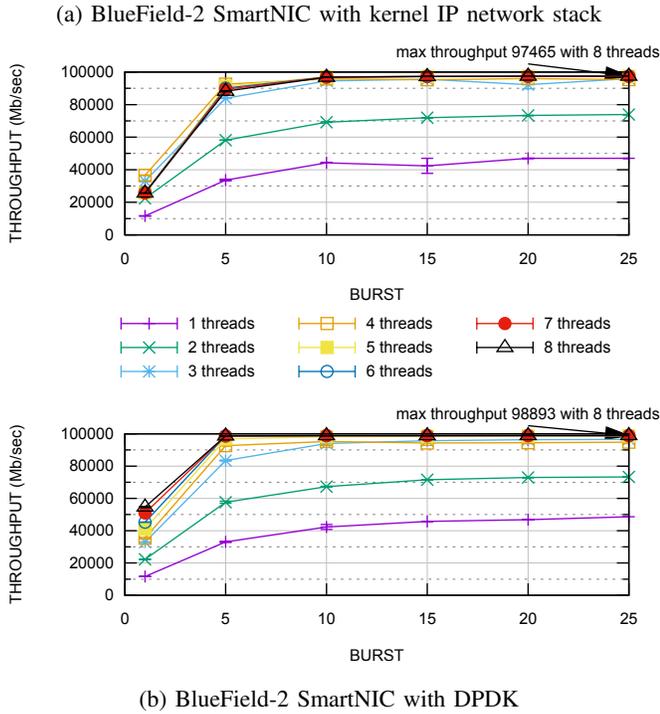

  \begin{subfigure}{\columnwidth}
    \caption{\bf2 SmartNIC with kernel IP network stack}
    \label{fig:r7525_pkt_size_832bytes_embedded_kernel}
    \includesvg[inkscapelatex=false, width=\columnwidth]{images/r7525_pkt_size_832bytes_embedded_kernel.svg}
  \end{subfigure}
  \begin{subfigure}{\columnwidth}
    \includesvg[inkscapelatex=false, width=\columnwidth]{images/r7525_pkt_size_832bytes_embedded_dpdk.svg}
    \caption{\bf2 SmartNIC with DPDK}
    \label{fig:r7525_pkt_size_832bytes_embedded_dpdk}
  \end{subfigure}
  \caption{Throughput results from the r7525 machine in embedded function mode (packet size 832B)}
  \label{fig:pktgen_r7552_host_embedded_function_mode}
\end{figure}

\begin{figure}[!ht]
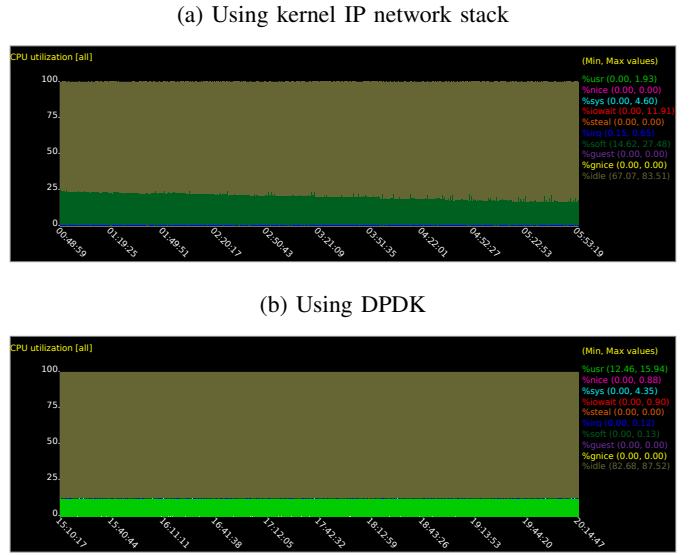

  \begin{subfigure}{\columnwidth}
    \caption{Using kernel IP network stack}
    \label{fig:pkt_size_832bytes_MBF_embedded_kernel_sys_activity}
    \includesvg[inkscapelatex=false, width=\columnwidth]{images/pkt_size_832bytes_MBF_embedded_kernel_sys_activity.svg}
  \end{subfigure} \par\bigskip
  \begin{subfigure}{\columnwidth}
    \caption{Using DPDK}
    \label{fig:pkt_size_832bytes_MBF_embedded_dpdk_sys_activity}
    \includesvg[inkscapelatex=false, width=\columnwidth]{images/pkt_size_832bytes_MBF_embedded_dpdk_sys_activity.svg}
  \end{subfigure}
  \caption{\bf2 SmartNIC CPU utilization in embedded function mode}
  \label{fig:smartnic_embedded_function_mode_cpu_utilization}
\end{figure}

\subsubsection{Evaluation in the Embedded Function Mode}
For completeness, we conducted additional experiments where the \bf2 was configured to run in the embedded function mode. As discussed in Section~\ref{sec:bf2_spec}, this mode is typically used in network security applications where packets must be inspected and manipulated in real time. The data path for this mode places the ARM processor \emph{between} the network and the host. There are two main benefits to this architecture. First, the host's workload should decrease because packet filtering operations and network data structures are offloaded to the SmartNIC. Second, transfer performance between the SmartNIC and the host should improve because the \bf2 supports DPDK~\cite{b53}. DPDK is an open source library that includes a number of performance optimizations for transferring packets between user-space applications and hardware.

As a means of observing how much processing headroom is available on the SmartNIC when running in the embedded function mode, we configured the host to generate a stream of packets and then measured both the throughput of the host and the CPU load of the ARM. As illustrated in Figure~\ref{fig:r7525_pkt_size_832bytes_embedded_kernel}, the host was able to saturate the network link with packets as small as 832B. The ARM processors in this configuration simply transferred packets from card memory to the network without any processing. As such the ARM cores were observed to have 75\% to 82\% of their CPU load available for other computations. To evaluate the overhead of the kernel IP network stack, we ran the same test but configured the \bf2 to use DPDK. The efficiency of the DPDK implementation gives an additional boost to the CPU availability for applications, translating to 5.5\% to 12.5\% more CPU time than the kernel network solution on the \bf2 (Figure~\ref{fig:pkt_size_832bytes_MBF_embedded_dpdk_sys_activity}).

These results indicate that in Ethernet networking scenarios where the separated host mode is not specifically required, the embedded function mode may be a much more desirable mechanism for implementing in-transit computations. Additionally, user space network stacks such as DPDK may offer significant improvements.

\section{Profitable Types of Operations for Offloading}
\label{sec:profitable_operations}

\subsection{Benchmark Considerations}
The goal of our second performance evaluation is to characterize the \emph{computational} strengths and weaknesses of the \bf2 hardware. This characterization is important because it helps us determine scenarios where it may be profitable to offload computations to SmartNICs that are in a workflow's data path. As demonstrated in the previous section, it is often difficult to determine which hardware components in an embedded device will have the most influence on performance without a great deal of experimentation. Therefore this evaluation focuses on running a wide range of microbenchmarks that help illuminate many different aspects of the hardware's capabilities. 

The \stressng tool was selected for this evaluation because it contains a large number diverse thrashing functions (called ``stressors'') that are designed to stress different software and hardware components of a system. In contrast to integrated tests, each \stressng stressor repeats a specific operation continuously for a fixed period of time. For example, the msync stressor tests the msync(2) system call while the CPU stressor tests the CPU's floating-point, integer, and bit manipulation performance separately, each with a different function. \stressng contains a total of 250 stressors that cover a wide range of resource and function domains, including disk IO, network IO, DRAM, filesystem, system calls, CPU operations, and CPU cache. Inside \stressng these domains are used as classes to categorize stressors. Our evaluation collected performance measurements for several systems and analyzes differences at both the individual stressor level as well as the broader class levels.

One challenge in comparing the performance of different stressors is that results are reported in terms of ``bogus operations per second'' or bogo-ops-per-second. Each stressor simply counts the number of times a particular operation can be performed in a given amount of time. While this metric provides a means of comparing how well \emph{different systems} perform the same task, it is meaningless to directly compare the bogo-ops-per-second numbers of \emph{different stressors}. Listing~\ref{listing:stress-ng_example_results} shows example results for two stressors that executed on the same machine. 

\begin{listing}[!ht]
\begin{minted}{yaml}
- stressor: branch
  bogo-ops: 14511254875
  bogo-ops-per-second-usr-sys-time: 36305366.212159
  bogo-ops-per-second-real-time: 1451159699.206867
  wall-clock-time: 9.999764
  user-time: 399.700000
  system-time: 0.000000

- stressor: memrate
  bogo-ops: 283
  bogo-ops-per-second-usr-sys-time: 0.700877
  bogo-ops-per-second-real-time: 27.980803
  wall-clock-time: 10.114077
  user-time: 388.940000
  system-time: 14.840000
\end{minted}
  \caption{Stress-ng Example Results}
  \label{listing:stress-ng_example_results}
\end{listing}

\subsection{Experiment Setup and Results}
The \stressng test was run on the \bf2 SmartNIC as well as a variety of host systems available in CloudLab to better understand the computational capabilities of the card. As listed in Table~\ref{table:stress-ng_test_platforms} the 12 host systems all used Intel x86\_64 processors with the exception of the m400 system, which is based on the older ARMv8 A57. While it may not be fair to compare the \bf2 card's embedded processor to general-purpose server processors, all of the servers were at least three years old. The d710 is based on a 12-year old processor.

\begin{table*}[!ht]\scriptsize
\begin{threeparttable}
\caption{Specifications of Stress-ng Test Platforms}
\label{table:stress-ng_test_platforms}
\begin{tabularx}{\linewidth}{ l l l l l l } \toprule
\textbf{Platform} & \textbf{CPU (Release Date)} & \textbf{Cores} & \textbf{DRAM} & \textbf{Disk} & \textbf{NIC} \\ \toprule
c220g1  & Intel E5-2630 v3 @ 2.4 GHz (Q3'14) & 2 x 8 & 128GB (DDR4-1866) & 2 x SAS HDD, 1 x SATA SSD & 2 x 10Gb, 1 x 1Gb \\
c220g2  & Intel E5-2660 v3 @ 2.6 GHz (Q3'14) & 2 x 10 & 160GB (DDR4-2133) & 2 x SAS HDD, 1 x SATA SSD & 2 x 10Gb, 1 x 1Gb \\
c220g5  & Intel Xeon Silver 4114 @ 2.2 GHz (Q3'17) & 2 x 10 & 192GB (DDR4-2666) & 1 x SAS HDD, 1 x SATA SSD & 2 x 10Gb, 1 x 1Gb \\
c6220   & Xeon E5-2650 v2 @ 2.6 GHz (Q3'13) & 2 x 8 & 64GB (DDR3-1866) & 2 x SATA HDD & 2 x 10Gb, 4 x 1Gb \\
c8220   & Intel E5-2660 v2 @ 2.2 GHz (Q3'13) & 2 x 10 & 256GB (DDR3-1600) & 2 x SATA HDD & 2 x 10Gb, 1 x 40Gb IB \\
d430    & Intel E5-2630 v3 @ 2.4 GHz (Q3'14) & 2 x 8 & 64GB (DDR4-2133) & 2 x SATA HDD, 1 x SATA SSD & 2 x 10Gb, 2 x 1Gb \\
d710    & Intel Xeon E5530 @ 2.4 GHz (Q1'09) & 1 x 4 & 12GB (DDR3-1066) & 2 x SATA HDD & 4 x 1Gb \\
dss7500 & Intel E5-2620 v3 @ 2.4 GHz (Q3'14) & 2 x 6 & 128GB (DDR4-2133) & 45 x SATA HDD, 2 x SATA SSD & 2 x 10Gb \\
m400    & ARMv8 Atlas/A57 (64-bit) @ 2.4 GHz & 1 x 8 & 64GB (DDR3-1600) & 1 x M.2 SSD & 2 x 10Gb \\
m510    & Intel Xeon D-1548 @ 2.0 GHz (Q4'15) & 1 x 8 & 64GB (DDR4-2133) & 1 x NVMe SSD & 2 x 10Gb \\
r320    & Xeon E5-2450 @ 2.1 GHz (Q2'12) & 1 x 8 & 16GB (DDR3-1600) & 4 x SATA HDD & 2 x 1Gb \\
xl170   & Intel E5-2640 v4 @ 2.4 GHz (Q1'16) & 1 x 10 & 64GB (DDR4-2400) & 1 x SATA SSD & 4 x 25Gb \\
MBF2H516A-CENO\_Ax & ARMv8 A72 (64-bit) @ 2.5 GHz & 1 x 8 & 16GB (DDR4-1600) & eMMC flash memory & 2 x 100 Gb/s or 1 x 200Gb/s \\
\bottomrule
\end{tabularx}
\begin{tablenotes}\scriptsize
\item 1. All platforms except the MBF2H516A-CENO\_Ax (\bf2 SmartNIC) ran Ubuntu 20.04 (kernel 5.4.0-51-generic).
\item 2. Tests were all conducted on the ext3 filesystem.
\end{tablenotes}
\end{threeparttable}
\end{table*}

In addition to the CloudLab systems, we conducted the \stressng experiments on a stock Raspberry Pi 4B (RPi4) system with 4GB of RAM. While the RPi4 is not a particularly fast embedded system, it is ubiquitous and serves as a reference hardware platform that others may use for comparing results. Performance numbers reported in this section are normalized to the RPi4. 

\begin{figure*}[!ht]
  \includesvg[inkscapelatex=false, width=\textwidth]{images/sequential-all_60s_Npc3000.platforms_summary.svg}
  \caption{Box plotting the relative performance of different stress-ng stressors for 12 general-purpose platforms and the SmartNIC. The run time of each stressor was 60 seconds. We used the 4 GB model of the Raspberry Pi 4B as the reference platform for performance normalization. The MBF2H516A-CENO\_Ax platform is the model name of the \bf2 SmartNIC in question. The data points of the SmartNIC are marked with triangles. The data points of the other platforms are plotted only if they are outliers (outside of the range of the corresponding whisker). Stressors without any data points are because they are not executed, hence they remain empty in the figure (e.g., aio and ioport).}
  \label{fig:sequential-all_60s_Npc3000.platforms_summary}
\end{figure*}

On each platform, we sequentially ran all default stressors (a total of 218\footnote{Some stressors were not executed on any platform because we did not use root privileges when running the tests.}), each of which ran for 60 seconds. Each stressor launches one instance on each online CPU core. We repeated this execution on a platform five times and averaged the results. To calculate the relative performance of a stressor for a platform, we divide the mean bogo-ops-per-second value for the platform by the corresponding value obtained for the RPi4. Some stressors did not run on some platforms because of a lack of support for a required capability. For example, the rdrand stressor did not run on the \bf2 SmartNIC as well as on the m400 platform because the ARM CPU does not support the rdrand instruction. The final results from all platforms are plotted in Figure~\ref{fig:sequential-all_60s_Npc3000.platforms_summary}.

\subsubsection{Individual Results Analysis}
As we expected, the \bf2 card's performance generally ranked lowest of all the systems tested except the RPi4. However, there were multiple tests where the \bf2 excelled and warranted a closer inspection, summarized below. The number in parentheses is the performance ranking of the \bf2 for a given stressor among all test platforms. \vspace{2mm} \\
\textbf{af-alg (\#1):} AF\_ALG~\cite{b17,b54} is the kernel crypto API user space interface. It allows user programs to access the cipher and hash functions provided by hardware cryptographic accelerators. The \bf2 SmartNIC contains multiple hardware accelerators for cryptography such as the IPsec/TLS data-in-motion accelerator, the AES-XTS 256/512-bit data-at-rest accelerator, the SHA 256-bit accelerator, and the true random number accelerator. These all contribute to the outstanding performance on this stressor test.
\vspace{2mm} \\
\textbf{lockbus and mcontend (\#1):} The lockbus stressor keeps step forwarding a pointer while injecting write barriers in between. The mcontend stressor starts a couple of threads to concurrently update and read data residing in memory that is mapped to the same physical page. These two stressors simulate aggressive memory access patterns that could be more extensive than real-world programs. However, the \bf2 handles these memory access contentions very well.
\vspace{2mm} \\
\textbf{stack (\#1), mremap (\#3), stackmmap and madvise (\#5), msync (\#6), mmap (\#8), malloc, and vm (\#13):} These stressors exercise the virtual memory subsystem of the OS running on the SmartNIC. Although we see the performance of accessing some memory interfaces on the \bf2 is significantly better than other servers, the most common memory interfaces, such as mmap and malloc, do not perform well on this device.
\vspace{2mm} \\
\textbf{chattr and inode-flags (\#5), ioprio (\#6), file-ioctl (\#7), dnotify and getdent (\#12), copy-file, dentry, dir, and fstat (\#13): } These stressors touch different interfaces provided by the filesystem. We see a considerable variation among these tests. It is important to note that the \bf2 does not show a performance advantage over other platforms when accessing other common interfaces for metadata operations, such as getdent(2) and fstat(2).
\vspace{2mm} \\
\textbf{fp-error (\#7), vecmath (\#9), branch, funccall, bsearch, hsearch, lsearch, qsort and skiplist (\#11), longjmp and shellsort (\#12), cpu, opcode, and tsearch (\#13):} These stressors focus on logic and arithmetic operations and include a variety of sorting tests. The cpu stressor performs a variety of math functions (e.g., manipulating bits, taking square roots, finding the greatest common divisor, and  computing Apéry's constant). Surprisingly, the relative performance of the \bf2 on this stressor is less than 1, meaning its arithmetic performance is \emph{even worse} than the RPi4. In contrast, the performance result of the vecmath stressor is interesting because the \bf2 performs \emph{better} than some of the x86\_64 CPUs on this test, including the D-1548 (Q4'15), the E5-2450 (Q2'12), and the Xeon E5530 (Q1'09).
\vspace{2mm} \\
\textbf{cache (\#11), icache (\#13):} The cache stressor thrashes the last-level cache of the CPU by reading/writing content from/to the cache as fast as possible. In this test the \bf2 performs slightly better than the E5-2630 v3 (Q3'14), the E5-2660 v2 (Q3'13), and the Xeon E5-2450 (Q2'12). However, all are still worse than the RPi4. The reason for this discrepancy is that the last-level cache on the \bf2 is the L3 cache, while the last-level cache on the RPi4 is the L2 cache. The icache stressor tests the instruction cache performance of the CPU by generating load misses using different sizes of memory pages. Overall, the CPU cache on the \bf2 does not have competitive performance over other platforms.
\vspace{2mm} \\
\textbf{sigsegv (\#9), timerfd (\#10), signal, clock, and timer (\#11), itimer, sigpipe, sigsuspend, and sleep (\#12), nanosleep (\#13):} The performance of these stressors represents the interrupt performance of the OS. These results indicate that software offloaded to the \bf2 should avoid using the OS's timing and interrupt interfaces when possible.
\vspace{2mm} \\
\textbf{readahead (\#11), hdd and seek (\#13):} The local storage hardware of the \bf2 is an eMMC flash device, while the CloudLab general-purpose servers have either enterprise-class HDDs or SSDs. The relative IO performance of the \bf2 is worse than most of the other platforms and suggests that offloaded functions should minimize their access to local storage.
\vspace{2mm} \\
\textbf{sem-sysv (\#2), fifo (\#9), eventfd, poll (\#11), futex (\#12), hrtimers (\#12), clone, exec, fork, nice, and pthread (\#13):} System V inter-process communication (IPC)~\cite{b55} is a set of mechanisms provided by Linux to simplify the communication between processes. The sem-sysv stressor measures how fast a pair of processes can safely increment and decrement a shared semaphore when mixing both legal and illegal arguments. The \bf2 SmartNIC's performance was better than all x86\_64 platforms in this test. This result may be due to an architectural advantage of the ARM processors, as the m400 ARM and RPi platforms also ranked higher than expected. However, other scheduler-related tasks did not perform well on the \bf2. For example, the \bf2 scored poorly on the futex stressor, which uses the futex(2)~\cite{b56} system call to wait until a condition becomes true.
\vspace{2mm} \\
\textbf{sockabuse (\#11), epoll, sockmany, sock, udp-flood, and udp (\#13):} These stressors test the performance of the kernel network stack. The rankings of these stressors show the networking performance of the \bf2 using the kernel stack is worse than most of the other platforms in the comparison. This result is consistent with our previous findings from evaluating the processing headroom under packet transmission.
\vspace{2mm}

We list the performance rankings of the SmartNIC among all test platforms for each stress test in Table~\ref{table:bluefield-2_rankings_based_on_stressor_tests}. Considering the large number of stressors, we believe that the most useful way to represent this information is to show the best and worst results for the \bf2. This ordering provides us with guidance on the types of operations that should and should not be performed on the \bf2.

\begin{table}[!ht]
\begin{threeparttable}
\caption{Performance ranking of the \bf2 SmartNIC based on the results of stressor tests}
\label{table:bluefield-2_rankings_based_on_stressor_tests}
\begin{tabularx}{\linewidth}{ l l l } \toprule
\textbf{Stressor} & \textbf{Stressor Classes} & \textbf{Ranking} \\ \toprule
af-alg & CPU $|$ OS   &    1 \\
klog & OS   &    1 \\
lockbus & CPU\_CACHE $|$ MEMORY  &  1 \\
mcontend & MEMORY   &    1 \\
splice & PIPE\_IO $|$ OS &  1 \\
stack & VM $|$ MEMORY  &   1 \\
dev & DEV $|$ OS & 2 \\
sem-sysv & OS $|$ SCHEDULER   &    2 \\
get & OS    &    3 \\
mremap & VM $|$ OS    &    3 \\
chattr & FILESYSTEM $|$ OS    &    5 \\
inode-flags & OS $|$ FILESYSTEM  & 5 \\
madvise & VM $|$ OS   &    5  \\
personality & OS    &    5 \\
stackmmap & VM $|$ MEMORY & 5 \\
sysinfo & OS  &  5 \\
ioprio & FILESYSTEM $|$ OS    &    6 \\
msync & VM $|$ OS & 6 \\
brk & OS $|$ VM  &  7 \\
file-ioctl & FILESYSTEM $|$ OS  &  7 \\
fp-error & CPU & 7 \\
bigheap & OS $|$ VM   &    8 \\
mknod & FILESYSTEM $|$ OS & 8 \\
mmap & VM $|$ OS  &  8 \\
revio & IO $|$ OS &  8 \\
context & MEMORY $|$ CPU  &  9 \\
dirdeep & FILESYSTEM $|$ OS   &    9 \\
fifo & PIPE\_IO $|$ OS $|$ SCHEDULER & 9 \\
locka & FILESYSTEM $|$ OS & 9 \\
lockofd & FILESYSTEM $|$ OS   &    9 \\
sigsegv & INTERRUPT $|$ OS    &    9 \\
vecmath & CPU $|$ CPU\_CACHE   &    9 \\
chown & FILESYSTEM $|$ OS & 10 \\
env & OS $|$ VM  & 10 \\
timerfd & INTERRUPT $|$ OS    &    10 \\
\midrule
bad-altstack & VM $|$ MEMORY $|$ OS & 14 \\
getrandom & OS $|$ CPU  &  14 \\
inotify & FILESYSTEM $|$ SCHEDULER $|$ OS  &  14 \\
netdev & NETWORK    &    14 \\
rename & FILESYSTEM $|$ OS  &   14 \\
resources & MEMORY $|$ OS & 14 \\
rseq & CPU   &   14 \\
schedpolicy & INTERRUPT $|$ SCHEDULER $|$ OS    &    14 \\
sigabrt & INTERRUPT $|$ OS    &    14 \\
sigchld & INTERRUPT $|$ OS    &    14 \\
vforkmany & SCHEDULER $|$ OS  &    14 \\
vm-addr & VM $|$ MEMORY $|$ OS  &   14 \\
\bottomrule
\end{tabularx}
\begin{tablenotes}\scriptsize
\item * We only show the stressors that \bf2 SmartNIC ranks $\leqslant$ 10 or the last among all test platforms.
\end{tablenotes}
\end{threeparttable}
\end{table}

As a means of determining whether thermal or caching effects were impacting performance, we reduced the duration of each stressor from 60 seconds to 10 seconds and repeated the tests on all platforms. Table~\ref{table:bluefield-2_ranking_changes} lists the stressors where the \bf2 changed more than two positions in the overall rankings. The stressors with the biggest change are the CPU and CPU cache-related stressors. For example, the bigheap stressor exercises the virtual memory by bumping up the memory of a process with the REALLOC(3)~\cite{b19} system call until an out-of-memory error is triggered. Furthermore, the rankings in the 60 second test are mostly higher than the stressors' corresponding rankings in the 10 second test. This information suggests that the ARM CPU on the \bf2 may need to be warmed up for optimal performance. Regarding the thermal dissipation, we have not seen a noticeable impact it caused on the performance of the \bf2.

\begin{table}[!ht]
\begin{threeparttable}
\caption{Changes in the performance ranking of the \bf2 SmartNIC in the 10s and 60s tests}
\label{table:bluefield-2_ranking_changes}
\begin{tabularx}{\linewidth}{ l l l l } \toprule
\textbf{Stressor} & \textbf{Stressor Classes} & \textbf{10s Test} & \textbf{60s Test} \\ \toprule
af-alg & CPU $|$ OS    &    7 & 1 \\
bigheap & OS $|$ VM & 14 & 8 \\
branch & CPU    & 14 & 11 \\
brk & OS $|$ VM & 11 & 7 \\
cache & CPU\_CACHE   &  14 & 11 \\
dirdeep & FILESYSTEM $|$ OS & 13 & 9 \\
klog & OS   &   5 & 1 \\
seek & IO $|$ OS & 7 & 13 \\
sigfd & INTERRUPT $|$ OS & 14 & 11 \\
\bottomrule
\end{tabularx}
\end{threeparttable}
\end{table}

\subsubsection{Class Results Analysis}
\begin{sidewaysfigure*}[!ht]
  \includesvg[inkscapelatex=false, width=\textwidth]{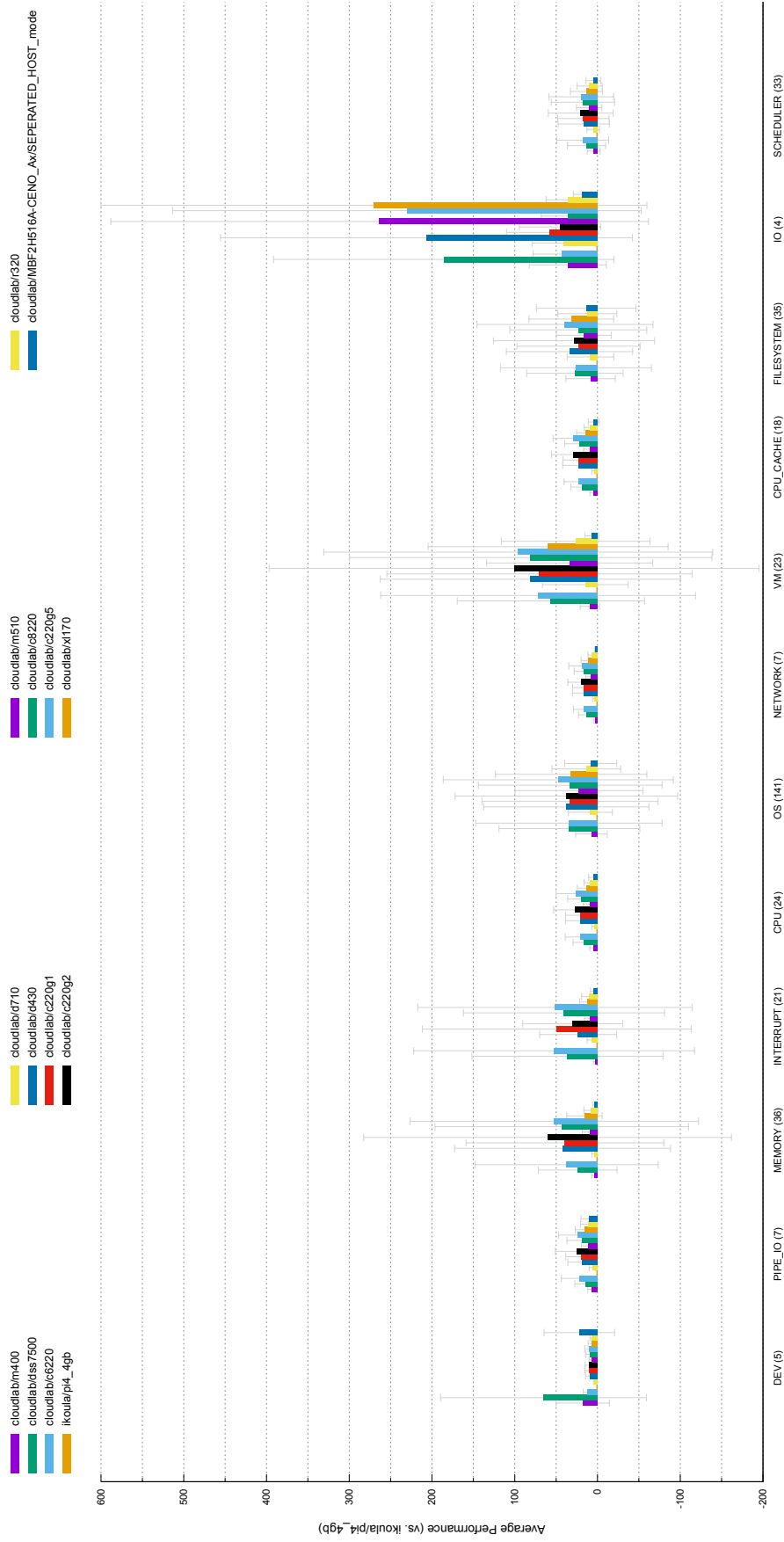}
  \caption{The average relative performance of the stressors in a stressor class for a particular platform. If a stressor belongs to multiple classes, its relative performance value will be added to each of the belonging classes in the calculation. The number beside the name of a stressor class (along the x-axis) is the number of stressors in that class. The whisker on a bar is the sample standard deviation of the average. The MBF2H516A-CENO\_Ax platform is the \bf2 SmartNIC in question.}
  \label{fig:sequential-all_60s_Npc3000.platforms_summary.per_class}
\end{sidewaysfigure*}

\stressng categorizes all stressors into 12 classes. To evaluate whether the \bf2 has performance advantages over other platforms in certain classes of operations, we calculate the average relative performance of the stressors in a stressor class for each class and each platform and show the result in Figure~\ref{fig:sequential-all_60s_Npc3000.platforms_summary.per_class}. The relevant class definitions based on our understanding are listed below:

\begin{itemize}
    \item \texttt{DEV:} Stressors that test a series of Linux device interfaces under /dev, e.g., /dev/mem, /dev/port, /dev/null, /dev/loop*.
    \item \texttt{PIPE\_IO}: Stressors that test Linux pipe I/O interfaces, e.g., fifo, sendfile(2)~\cite{b21}, tee(2)~\cite{b22}, and splice(2)~\cite{b23}.
    \item \texttt{MEMORY}: Stressors that test the memory subsystem of a machine. Example stressors are malloc, memrate, stackmmap, and mcontend.
    \item \texttt{INTERRUPT}: Stressors that test different Linux interrupt signals. Example stressors are timerfd, sigrt, timer, and sigq.
    \item \texttt{CPU}: Stressors that thrash different functionalities of the CPU such as encryption, atomic operations, random number generation, and arithmetic calculation. Example stressors are crypt, atomic, getrandom, cpu, nop, matrix-3d, and vecmath.
    \item \texttt{OS}: Stressors that test various general system interfaces. Typically, stressors of this class also belong to some other classes. For example, the aforementioned stressor fifo, sigrt, and malloc belong to this class as well.
    \item \texttt{NETWORK}: Stressors that test the network performance and functionality of the system. Example stressors are sockfd, sockmany, rawpkt, udp-flood, and sctp.
    \item \texttt{VM}: These stressors focus on testing the virtual memory layer managed by the operating system. Even though as a process running on Linux, it is hard to test the hardware memory subsystem without testing the virtual memory; stressors in this class focus on some high-level virtual memory operations such as mlock~\cite{b24}, msync~\cite{b25}, swapon/swapoff~\cite{b26}, and madvise~\cite{b27}.
    \item \texttt{CPU\_CACHE}: Stressors that can thrash the CPU cache. The previously discussed cache and icache stressors belong to this class.
    \item \texttt{FILESYSTEM}: As the name indicates, these stressors test various filesystem interfaces. Example stressors are iomix, xattr, flock, getdent, and fallocate.
    \item \texttt{IO}: These stressors try to test the raw storage performance as heavily as possible. Example stressors are revio, hdd, sync-file, and readahead.
    \item \texttt{SCHEDULER}: These stressors test the capability and stability of the system scheduler. Example stressors are zombie, nice, affinity, mq, spawn, and yield.
\end{itemize}
A complete list of stressors and their associated classes can be found on~\cite{b51}.
\vspace{2mm}

From Figure~\ref{fig:sequential-all_60s_Npc3000.platforms_summary.per_class} we can see that the average performance of the \bf2 is on par with the 12-year-old x86\_64 server d710 and the ARM server m400. However, since the average performance of almost all stress classes has a large variation, the comparison between them becomes less meaningful because the differences are not statistically significant. For example, the IO stressor class only has four stressors, but no matter which test platform, the standard deviation of the stressor results is almost as large as the average performance. This figure signifies, in general, the SmartNIC platform is weaker than a general-purpose x86\_64 platform (except for pretty old ones) in most operations. There is not any class of operations that the \bf2 can perform statistically better than a traditional server. That means we can only discover the advantages of the \bf2 by looking at the performance of individual operations.

\subsubsection{Summary and Insights}
To summarize our findings, the advantage boundary of the \bf2 is small. Therefore, offloading functions to the SmartNIC requires careful consideration about the resource usage and operations of the functions to ensure the design is tailored to the given embedded environment. Specifically, the function should avoid containing operations working on local storage and filesystem IO, heavy CPU calculation, or relying on frequent software interruptions and system scheduling. Most importantly, avoid using the kernel network stack whenever possible by migrating the network stack to either user space or hardware-accelerated solutions. With that being said, some operations are profitable to be offloaded to the SmartNIC such as memory contention operations, cryptographic operations, and IPC operations. In general, executing memory-related operations is better than executing CPU, storage IO, and kernel network-related operations on the SmartNIC. The exceptions to the CPU operations are encryption and vector calculations. The former is due to the built-in accelerators in the SmartNIC. The latter may be due to the optimization of the ARM CPU.

Based on the analysis, one type of function that is potentially profitable to be offloaded to the SmartNIC is transparent encryption/decryption or compression/decompression for the data sent/received to/from the host. Moving this function to the SmartNIC can significantly save CPU cycles for applications running on the host, and at the same time, help reduce the function execution latency, according to our evaluation results. Other types of functions that may also be profitable for offloading are those that can adequately take advantage of the efficiency in (virtual) memory access operations, IPC operations, and/or vector calculations of the SmartNIC. A concrete example is the data transformation function using Apache Arrow~\cite{b48} -- the de facto in-memory data representation library. This library uses vectors as the basic unit to store data in a columnar format. It also provides an IPC mechanism to synchronously transfer collections of Arrow columnar arrays between processes. In such a way, data transformation between different programs can use the well-defined data representation to bridge the two ends without needing to translate between on-disk and in-memory representations. Apache Arrow has developed a large open-source community that attracts data scientists and software developers to build Arrow-based data processing systems~\cite{b46,b47,b49,b18}. The Magpie system~\cite{b45} from Microsoft is one of them.

\section{Conclusion} \label{sec:conclusion}
In this paper, we evaluated the maximum processing headroom for packet transmission workloads running on the \bf2 SmartNIC, showing that the CPU time available for applications running inside of the SmartNIC is limited unless the SmartNIC's embedded environment is appropriately configured and avoids the use of the kernel IP stack for networking. Followed by that, we used \stressng to generate the performance profiles for different operations on multiple general-purpose servers, and compared them with the ones obtained from the SmartNIC. The results show that the performance capability of the SmartNIC is in general behind that of servers in the current data center. However, three types of operations tend to be relatively profitable to be offloaded to the SmartNIC to acquire better performance, including data encryption operations, memory operations that involve heavy contention, and IPC operations.

\end{document}